\documentclass[preprint, 12pt,nofootinbib]{revtex4-1}
\usepackage{graphicx}
\usepackage{cancel}
\usepackage{textcomp}
\usepackage{amsmath, amssymb, amsfonts, latexsym, epsfig}
\usepackage{mathrsfs}

\usepackage{bm}
\usepackage{times}
\usepackage{epsfig}
\usepackage{color}

\def\beq{\begin{equation}}
\def\eeq{\end{equation}}
\def\be{\begin{equation}}
\def\ee{\end{equation}}
\def\bea{\begin{eqnarray}}
\def\eea{\end{eqnarray}}
\def\to{\rightarrow}

\begin{document}
\title{What if $b\bar{b}$ does not dominate the decay of the Higgs-like boson?}
\author{Jiwei Ke}
\author{Hui Luo}
\author{Ming-xing Luo}
\author{Tian-yang Shen}
\author{Kai Wang}
\author{Liucheng Wang}
\author{Guohuai Zhu}
\affiliation{ Zhejiang Institute of Modern Physics and Department of Physics, Zhejiang University, Hangzhou, Zhejiang 310027, CHINA}

\begin{abstract}
{
The dominant decay mode of standard model Higgs at 126~GeV $b\bar{b}$
suffers from severe SM background at the LHC even in associated productions $Wh_{\rm SM}$ or $Z h_{\rm SM}$. The precision measurement of  BR($\phi\to b\bar{b})$ requires more data to reduce its large error bar. We investigate the possibility of this channel with largest uncertainty not dominating the decay of Higgs-like boson discovered at the LHC. In such scenarios, the Higgs signal shows
highly suppressed $b\bar{b}$, slightly reduced $\tau^{+}\tau^{-}$ and moderately enhanced gauge bosons comparing with the SM predictions. The model requires two different sources of electroweak symmetry breaking and radiative correction to $m_{b}$ strongly enhanced. However, large reduction in $b\bar{b}$ usually results large enhancement in $\tau^{+}\tau^{-}$ mode in particular. The reduction of $\tau^{+}\tau^{-}$ therefore implies that a new decay mode is inevitable. We find that a non-decoupling MSSM Higgs decay into lighter Higgs $H\to hh$ may fit the signature. Here, MSSM $H$ is identified as the 126~GeV resonance while $h$ is below $M_{H}/2$ and can evade the direct search bound at LEPII and Tevatron. Large $PQ$ and $R$ symmetry breaking effects mediated by strong interaction can strongly enhance radiative corrections in $m_b$. However, the
scenario can only be realized in highly fine-tuned parameter region where $G_{Hhh}$ is tiny. Nevertheless, we discuss the discovery potential of this highly fine-tuned $H\to hh$ at the LHC. }
\end{abstract}

\maketitle

The ATLAS and CMS collaborations at the CERN Large Hadron Collider (LHC) has
discovered a Higgs-like boson $\phi$ of 126~GeV via various channels.  It was first seen via the two cleanest channels, the di-photon ($gg\to \phi\to \gamma\gamma$) and the four-lepton ($gg\to \phi\to ZZ^{*}\to \ell^{+}_{i}\ell^{-}_{i}\ell^{+}_{j}\ell^{-}_{j}$ with $i,j=e^{\pm},\mu^{\pm}$) modes\cite{today} and later in the di-lepton  ($gg\to \phi\to WW^{*}\to \ell^{+}_{i}\nu_{i}\ell^{-}_{j}\bar{\nu}_{j}$) with mass range consistent with the four-lepton measurement \cite{ww}. Both collaborations \cite{Atlas-Spin,CMS-Spin} recently also updated their studies on spin
and parity and a CP-even spin-zero state $J^{P}=0^{+}$ is preferred based on the data of four lepton channel. In addition, both collaborations have also reported the boson decaying into tau pairs, $\phi\to \tau^{+}\tau^{-}$. This is the first evidence at the LHC that the Higgs-like boson actually couples to SM fermions. On the other hand, for a SM Higgs boson of 126~GeV,
more data is required to reduce the large error bar in the dominant decay channel $h_{\rm SM}\to b\bar{b}$.
$h_{\rm SM}\to b\bar{b}$ from gluon fusion suffers tremendous QCD background so
it can only be searched through associated productions $W \phi$ and $Z\phi$ with leptonic
decays of $W/Z$. Due to large uncertainties in $b$-jet measurements, the reconstructed boson mass lies in a broad range.
There also exists large uncertainties in missing transverse energy $\cancel{E}_{T}$ measurements.
In addition, large gluon PDF  results in huge number of $b$-jets at the LHC which leads
signals of $W/Z+\phi\to b\bar{b}$ to suffer from severe background.
Both collaborations have seen enhancement in di-photon channels with respect
to the SM prediction and moderate reduction in the $\tau^{+}\tau^{-}$ channel. In the di-lepton and four-lepton
channels, results from two collaborations largely overlap. In the $b\bar{b}$ channel,
the central values are very different and the result of ATLAS collaboration still has very large uncertainty. In term of $R$, the ratio between observation and SM prediction $R\equiv \sigma_{\rm obs.}/\sigma_{\rm SM}$, latest results of $b\bar{b}$ from both collaborations are
\bea
{\rm ATLAS} &:&-0.4 \pm 1.1 (13~\text{fb}^{-1}@8~\text{TeV}, 4.7~\text{fb}^{-1}@7\text{TeV})\nonumber\\
{\rm CMS} & : & 1.1 \pm 0.6 (12.1~\text{fb}^{-1}@8~\text{TeV}, 5~\text{fb}^{-1}@7\text{TeV})
\label{r}
\eea
Given its large uncertainty,  it is worth investigating the possible scenario if $b\bar{b}$ channel does not dominate the Higgs decay. To illustrate the feature, we use  ATLAS central values to fit  other channels and assume that the $b\bar{b}$ is highly suppressed. The $R$-values are
\bea
R^{\rm ATLAS}_{\gamma\gamma} &=& 1.8\pm 0.5 (5.9~\text{fb}^{-1}@8~\text{TeV}, 4.8~\text{fb}^{-1}@7\text{TeV})\nonumber\\
R^{\rm ATLAS}_{4\ell} &=&1.4\pm 0.6 (5.8~\text{fb}^{-1}@8~\text{TeV}, 4.8~\text{fb}^{-1}@7\text{TeV})\nonumber\\
R^{\rm ATLAS}_{2\ell 2\nu}&=& 1.5 \pm 0.6(13~\text{fb}^{-1}@8~\text{TeV})\nonumber\\
R^{\rm ATLAS}_{\tau^{+}\tau^{-}}&=& 0.7\pm 0.6 (13~\text{fb}^{-1}@8~\text{TeV}, 4.6~\text{fb}^{-1}@7\text{TeV})
\label{r2}
\eea
 In this paper, we study whether the scenario of highly suppressed $b\bar{b}$, slightly reduced $\tau^{+}\tau^{-}$ with moderately enhanced gauge boson pairs can be realized in simple models.

One important property of the SM Higgs $h_{\rm SM}$ is that it couples SM fermions with strengths proportional to their masses. At Hadron colliders, $\tau$-lepton and $b$-jets
are  two leading and best identifiable final states in Higgs decaying into SM fermions.
Therefore, the comparison between these viable modes play an important role to test whether the Higgs-like boson
is the SM Higgs. Within SM, the ratio $X$ between branching fraction in $\tau^{+}\tau^{-}$ and $b\bar{b}$ channels is
\beq
X\equiv \frac{\text{BR}(h_{\rm SM}\to \tau^{+}\tau^{-})}{\text{BR}(h_{\rm SM}\to b\bar{b})}= \frac{\Gamma(h_{\rm SM}\to \tau^{+}\tau^{-})}{\Gamma(h_{\rm SM}\to b\bar{b})}\simeq \frac{m^{2}_{\tau}}{N_{C}m^{2}_{b}K}~,\label{bbtautau}
\eeq
where $m_{b}$, $m_{\tau}$ are the bottom quark and tau lepton masses respectively, $N_{C}=3$ is the color factor. $K$ accounts for QCD corrections of Higgs decaying into light quark states which is typically $1/1.5\sim 1/2$ for Higgs mass of ${\cal O}(120~\text{GeV})$. For $M_{h_{\rm SM}}=126$~GeV, $X\sim 1/10$. In $SU(5)$ Grand Unification, $\tau$ and $b$ arise from the same multiplet $\bf 5^{c}$. The ratios $X$ in Eq. \ref{bbtautau} in models originated from $SU(5)$ are naturally similar to the SM value. To obtain the highly reduced width of $H\to b\bar{b}$, additional radiative corrections must reduce the tree-level Yukawa couplings and split $b$ and $\tau$.
In order for radiative corrections to reduce tree level couplings, a second sector for electroweak symmetry breaking must exist. In Type-II Two-Higgs-Doublet-Models (2HDM),  both $b$ and $\tau$ masses arise from $\langle H_{d}\rangle$ at tree level but there  exist contributions of $\langle H_{u}\rangle$-type  from the mixing term
$M^{2}_{12}H_{u} H_{d}$ which naturally leads reduction in Yukawa couplings.
In SM, QCD correction of $\Gamma(h_{\rm SM}\to b\bar{b})$
can reduce the Born value from pole mass by 35\% to 50\% and we expect similar
mechanism in new physics models. A particularly interesting mechanism
lies in supersymmetric models where radiative corrections through strong interaction
are related to breaking of two symmetries, Peccei-Quinn symmetry and $R$-symmetry and
both symmetries must be broken at ${\cal O}$(TeV). For instance, the mixing term $H_{u}H_{d}$ in Type-II 2HDM
exists in soft supersymmetry breaking Lagrangian as $B\mu$-term which breaks both PQ and $R$-symmetry.
Due to strong interaction and large gluino mass, radiative corrections to $m_{b}$ are typically significantly enhanced.
This feature is highly non-trivial in beyond SM theories \cite{Carena:1995bx}.

Supersymmetric models are naturally Type-II 2HDM due to
holomorphic condition of superpotential and cancelations for $[SU(2)_{L}]^{2}U(1)_{Y}$ and Witten anomalies.
The SM fermion masses arise at tree level in superpotential
\beq
W=y_u Q u^{c} H_{u} +y_{d} Q d^{c} H_{d} +y_{\ell }\ell_{L} e^{c} H_{d}+\mu H_u H_d
\eeq
where $y_{u}$, $y_{d}$ and $y_{\ell}$ are  tree level Yukawa couplings of up, down quarks and charged leptons respectively. $y_d$ and $y_{\ell}$ are  proportional to $\tan\beta \equiv v_{u}/v_{d}$.
On the other hand, $Q d^{c} \bar{H}_{u}$ or $\ell e^{c} \bar{H}_{u}$, which are forbidden by the holomorphic condition of superpotential, is invariant under the SM gauge symmetries and can be generated as radiative corrections. Besides the SM gauge symmetries, listed in Table 1 are charge assignments of the particles under two additional symmetries in supersymmetric theory, Peccei-Quinn (PQ) symmetry and $R$-symmetry \footnote{To explicitly determine the charge assignments, we use $SU(5)$ convention in the Table 1.}.
PQ-symmetry which forbids the bare-$H_{u}H_{d}$ term in superpotential is explicitly broken by the $\mu$-term and $B\mu$-term. $R$-symmetry corresponds to the chiral symmetry that protects gaugino masses from being generated in the supersymmetric limit. $R$-symmetry breaking terms in the soft supersymmetry breaking Lagrangian are gaugino masses, $A$-terms and $B\mu$-term.
\begin{table}[h]
\begin{center}
\begin{tabular*}{1.0\textwidth}{@{\extracolsep{\fill}} c || c c c   c c c c | c}
\hline
Field & $Q$ & $u^{c}$ & $e^{c}$ & $d^{c}$ & $\ell$ & $H_{u}$ & $H_{d}$ & $\theta$ \\
\hline\hline
$R$-charge &  ${1\over 5}$ & ${1\over 5}$ & ${1\over 5}$ & $3\over 5$ & ${3\over 5}$ & $4\over 5$ & $6\over 5$ & 1 \\
PQ   & 0 & 0 & 0 & -1 & -1 & 0 & 1 & 0 \\
\hline
\end{tabular*}
\label{ccc}
\caption{Charge assignment under $R$-symmetry and Peccei-Quinn symmetry.}
\end{center}
\end{table}

Using charge assignments in Table 1, one can substitute them into calculation of effective coupling $Q d^{c} \bar{H}_{u}$ as
\bea
R[Q d^{c} \bar{H}_{u}] : & {1\over 5}+{3\over 5} - {4\over 5} =0 \\
{\rm PQ}[Q d^{c} \bar{H}_{u}]: &  0+ (-1) +0 =-1~.
\label{pq}
\eea
These equations clearly show that $Q d^{c} \bar{H}_{u}$ breaks both $R$-symmetry and PQ symmetry.
Therefore, radiative corrections of $m_{b}$ or $m_{\tau}$ in supersymmetry
are proportional to production of $\mu$ and gaugino masses or $A$-term.

On the other hand, for the SM Higgs $h_{\rm SM}$  with $M_{h}=126$~GeV, $b\bar{b}$ dominates 60\% of the Higgs decay with $\Gamma(h_{\rm SM}\to b\bar{b})=2.6\times 10^{-3}~\text{GeV}$. Significant reduction in $\Gamma(h_{\rm SM}\to b\bar{b})$ then results in significant reduction in total width and enhance the $WW$/$ZZ$ and di-photon if no new decay channels exists.
In particular, if $y_{\tau}$ reduction is not as large as $y_{b}$, $\text{BR}(\phi\to \tau^{+}\tau^{-})$ can also be significantly enhanced. However, since Eq.\ref{r2} shows the $R_{\tau^{+}\tau^{-}}\sim 70\%$, a new decay mode is then inevitable.

Two immediate options that can evade the current searches are the invisible decay of Higgs and the Higgs decaying into lighter scalars. The first option may be connected to the Dark Matter of the model but it is strongly constrained by requiring the relic density not to be over-abundant.
For the second option, $h\to AA$ in NMSSM \cite{jack,guiyu} or $H\to hh$ \cite{higherorder}
in non-decoupling MSSM \cite{cp, Heinemeyer2011,Bottino,taohan,hagiwara,heinemeyer,Arbey,Belanger:2012tt,Drees,Bechtle:2012jw,kejiwei,Bechtle:2012jw} may provide a simple realization. In this paper, we focus on the $H\to hh$ possibility of non-decoupling MSSM. On the other hand, in large parameter space, once $H\to hh$ decay is open, $\Gamma(H\to hh)$ is usually much larger than the other channels and may completely
dominate the decay of $H$. Therefore, $\Gamma(H\to hh)$  needs to
be highly fine-tuned to be at the comparable level as the width of SM Higgs
decaying into bottom pair $\Gamma(h_{\rm SM}\to b\bar{b})$.
In non-decoupling MSSM first proposed by \cite{cp}, $H$ is identified as the resonance at 126~GeV and a much lighter $h$ can evade direct searches in LEPII and Tevatron experiments by suppressing $ZZh$ coupling and thus production of $Zh$.
To reduce the $ZZh$ coupling which is the vacuum expectation value ({\it vev}) of $h$, simple realization is to let $h$ be the $H_{d}$-like boson since
large $m_{t}$ naturally requires large $v_{u}$. Given $h$ is a mixture state as $-\sin\alpha ({\rm Re}~H_{d}) + \cos\alpha ({\rm Re}~H_{u})$, this scenario prefers $\sin\alpha\simeq -1$ with large $\tan\beta$ which suppresses the $v_{d}$. In the limit of large $\tan\beta$ as $\sin\beta\to 1$, $\sin\alpha\to -1$ gives the $g_{ZZh}=\sin(\beta-\alpha)$ approaches zero.
In large parameter region,  the partial width $\Gamma(H\to hh)$ are typically several orders of magnitude higher than $\Gamma(h_{\rm SM}\to b\bar{b})$ and consequently $H\to hh$ completely dominate the $H$ decay there.
Therefore, visible decay channels of  $\gamma\gamma$, $ZZ^{*}\to 4\ell$ and $WW^{*}\to 2\ell 2\nu$ require
that the $\Gamma(H\to hh)\sim \Gamma(h_{\rm SM}\to b\bar{b})\sim 2\times 10^{-3}$~GeV.

Our numerical analysis are performed with the help of {\it FeynHiggs 2.9.2}~\cite{feynhiggs} with {\it HiggsBounds 3.8.0}~\cite{higgsbounds} and {\it SUSY\_Flavor 2.01}~\cite{Crivellin:2012jv}. We require that
\begin{itemize}
\item $M_{H}: 125\pm 2$~GeV;
\item $R_{\gamma\gamma}=\sigma^{\gamma\gamma}_{\rm obs}/\sigma^{\gamma\gamma}_{\rm SM}: 1\sim 2$;
\item Combined direct search bounds from HiggsBound3.8.0;
\item BR$(B\to X_{s}\gamma)<5.5\times 10^{-4}$;
\item BR$(B_{s}\to \mu^{+}\mu^{-})<6\times 10^{-9}$~.
\end{itemize}
In {\it FeynHiggs}, Higgs boson masses are calculated to full two-loop. To illustrate the qualitative feature here, we use the leading one-loop expression with only contributions of top Yukawa couplings.
Radiative corrections to
the Higgs boson mass matrix elements are \cite{Carena:1995bx,physicsreport}
\beq
\Delta {\cal M}^{2}_{12}\sim \sim \Delta {\cal M}^{2}_{11}\sim 0
\eeq
and
\beq
\Delta {\cal M}^{2}_{22} \sim \epsilon =\frac{3 m^{4}_{t}}{2\pi^{2}v^{2}\sin^{2}\beta}\left[ \log{M^{2}_{\rm SUSY}\over \bar{m}^{2}_{t}}+\frac{X^{2}_{t}}{2 M^{2}_{\rm SUSY}}\left(1- \frac{X^{2}_{t}}{6 M^{2}_{\rm SUSY}}\right)
 \right]
\eeq
where $X_{t}=A_{t}-\mu\cot\beta$ and $M^{}_{\rm SUSY}=(M_{\tilde{t}_{1}}+M_{\tilde{t}_{2}})/2$.

With one loop correction , the mixing angle $\alpha$ can be obtained
\beq
{\tan 2\bar{\alpha}\over \tan 2\beta} =\frac{M^{2}_{A}+m^{2}_{Z}}{M^{2}_{A}-m^{2}_{Z}+\epsilon/\cos 2\beta}~.
\label{alpha}
\eeq
The trilinear coupling among neutral Higgs bosons $H hh$, in unit of $-i \frac{m^{2}_{Z}}{v}$, is given in
\beq
\lambda_{Hhh}= [(2\sin 2\alpha \sin(\beta+\alpha)-\cos 2\alpha \cos(\beta+\alpha)]~,
\label{hhh1}
\eeq
with one-loop correction
\beq
\Delta \lambda_{Hhh}=   3\frac{\epsilon}{m^{2}_{Z}}\frac{\sin\alpha}{\sin\beta}\cos^{2}\alpha~.
\label{hhh2}
\eeq

\begin{figure}[h]
\begin{center}
\includegraphics[scale=1,width=7cm]{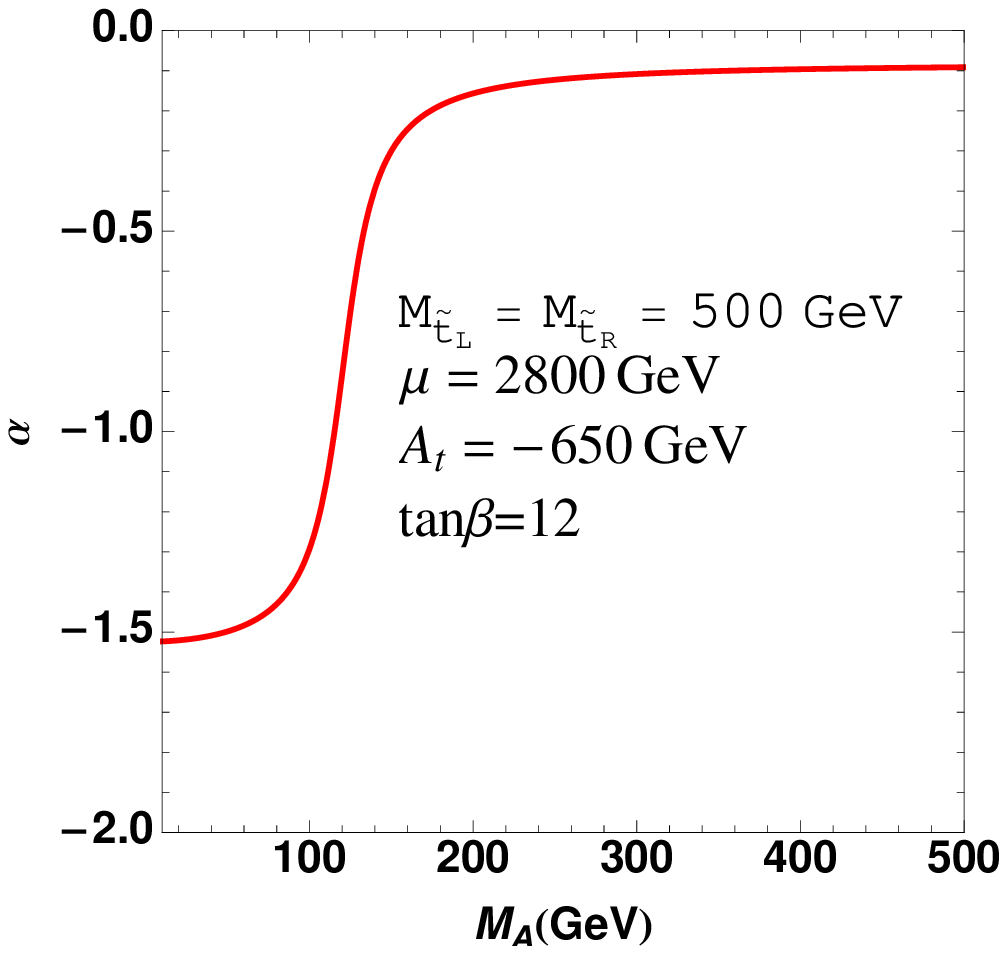}
\includegraphics[scale=1,width=7cm]{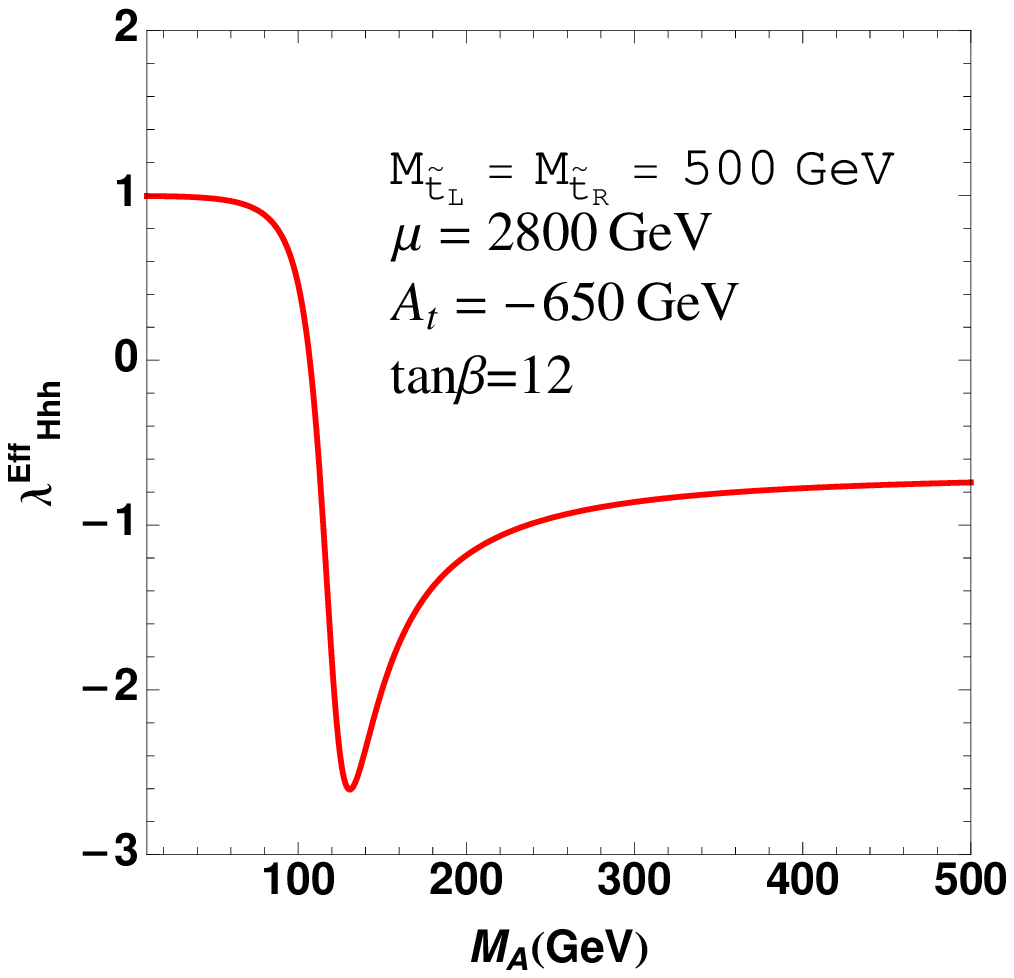}
\end{center}
\caption{(a) Mixing angle $\alpha$ and (b) normalized $Hhh$ coupling $\lambda_{Hhh}$ with respect to $M_{A}$, the other parameters
are fixed as shown explicitly in the plots.}
\label{mix}
\end{figure}

Plotted in Fig. \ref{mix} are the mixing angle $\alpha$ and normalized $Hhh$ coupling $\lambda_{Hhh}$ with respect to $M_{A}$ using one-loop result from Eqs \ref{alpha},~\ref{hhh1},~\ref{hhh2}  while the other parameters are fixed as in \cite{kejiwei}
\beq
\mu=2800~\text{GeV}, \tan\beta=12, M_{\tilde{t}_{L}}= M_{\tilde{t}_{R}}=500~\text{GeV}, A_{t}= -650~\text{GeV}~.
\eeq
Around $\alpha\sim -\pi/4$, $\lambda_{Hhh}$ vanishes.
Therefore, in order to get a highly reduced $\lambda_{Hhh}$, one can choose this fine-tuned region of $M_{A}$.

Explicitly the Yukawa couplings for $b$ and $\tau$ are \cite{Carena:1995bx}
\begin{equation}
\mathcal{L}=y_{b}H_{d}^{0}\bar{b}b+\triangle y_{b}H_{u}^{0}\bar{b}b+y_{\tau}H_{d}^{0}\bar{\tau}\tau+\triangle y_{\tau}H_{u}^{0}\bar{\tau}\tau,
\end{equation}
where $\Delta y_{i}$ stand for  corrections to the  Yukawa couplings and fermion masses arise from
\begin{equation}
m_{b}=y_{b}v_{d}+\triangle y_{b}v_{u},\qquad m_{\tau}=y_{\tau}v_{d}+\triangle y_{\tau}v_{u}.
\end{equation}
For simplicity, one can define
\begin{equation}
y_{b}=\frac{\sqrt{2}m_{b}}{v\cos\beta\left(1+\Delta_{b}\right)},\qquad y_{\tau}=\frac{\sqrt{2}m_{\tau}}{v\cos\beta\left(1+\Delta_{\tau}\right)}.
\end{equation}
where $\Delta_{b}=\triangle y_{b}\tan\beta/y_{b}$ ($\Delta_{\tau}=\triangle y_{\tau}\tan\beta/y_{\tau}$)
is the relative bottom (tau) mass correction.
In supersymmetric theory, leading contributions to $\Delta_{b}$ and $\Delta_{\tau}$ are
\begin{equation}
\Delta_{b}=\mu\tan\beta\left[\frac{g_{3}^{2}M_{3}}{6\pi^{2}}I\left(M_{3}^{2},m_{\tilde{b}_{1}}^{2},m_{\tilde{b}_{2}}^{2}\right)+\frac{y_{t}^{2}A_{t}}{16\pi^{2}}I\left(\mu^{2},m_{\tilde{t}_{1}}^{2},m_{\tilde{t}_{2}}^{2}\right)\right],\label{deltab}
\end{equation}
\begin{equation}
\Delta_{\tau}=\mu\tan\beta\left[\frac{g_{1}^{2}M_{1}}{16\pi^{2}}I\left(M_{1}^{2},m_{\tilde{\tau}_{1}}^{2},m_{\tilde{\tau}_{2}}^{2}\right)+\frac{g_{2}^{2}M_{2}}{16\pi^{2}}I\left(M_{2}^{2},\mu^{2},m_{\tilde{v}_{\tau}}^{2}\right)\right],\label{deltatau}
\end{equation}
where the positive-definite symmetric function $I$ is
\begin{equation}
I(x,y,z)=-\frac{xy\ln(x/y)+yz\ln(y/z)+zx\ln(z/x)}{(x-y)(y-z)(z-x)}.
\end{equation}
In Eq.\ref{deltab} and Eq.\ref{deltatau}, $\Delta_{b}$ and $\Delta_{\tau}$ are always proportional to the PQ-symmetry breaking term $\mu$.
However, $R$-symmetry breaking effects in $\Delta_{b}$ and $\Delta_{\tau}$ are very different. $\Delta_{b}$ is generated
through strongly interaction and enhanced by large gluino mass $M_{3}$. $\Delta_{\tau}$ only receives
correction via electroweak interaction with bino mass $M_{1}$. In addition, large $R$-symmetry breaking $y^{2}_{t}A_{t}$ also contributes.
As discussed in \cite{kejiwei}, squark loop may significantly cancel the contribution of light charged Higgs to flavor violation, in particular in $b\to s$ transition. Given the charged Higgs is
at similar scale as $M_{H}=126$~GeV in non-decoupling MSSM, scenario with
light top squark can survive all the flavor physics bounds. In $b\to s\gamma$, helicities for involved quark states must be flipped. This corresponds to a case with both chiral symmetry $U(3)_{Q}\times U(3)_{d}$ breaking and electroweak symmetry breaking and is exactly the same as the symmetry breaking in $m_{b}$ generation. The supersymmetric contribution to $b\to s\gamma$ is therefore exactly the same as supersymmetric correction to $m_{b}$. Large PQ-symmetry and $R$-symmetry breaking is also required to cancel $b\to s\gamma$ \cite{kejiwei}. The same parameter
choice can also improve the flavor violation in $B_{s}\to \mu^{+}\mu^{-}$.

We use one benchmark point to illustrate the feature discussed previously and discuss the collider phenomenology.
Without loss of generality, we fix masses of the  sfermions as \footnote{Scenarios with light stau \cite{carlos} can further enhance $R_{\gamma\gamma}$ but may reduce the $M_{H}$.}
\bea
M_{\tilde{Q}_{1,2}}&=&M_{\tilde{u}_{1,2}}=M_{\tilde{d}_{1,2,3}}=M_{\tilde{L}_{1,2,3}}=M_{\tilde{e}_{1,2,3}}=1~\text{TeV}~,\nonumber\\
M_{\tilde{Q}_{3}}&=&M_{\tilde{t}}=500~\text{GeV}
\eea
and gauginos as
\beq
M_{1}=200~\text{GeV}, M_{2}=400~\text{GeV}, M_{3}=1200~\text{GeV}~.
\eeq
Other parameters of benchmark point are listed as
\beq
 \mu = 2800~\text{GeV}, A_{t} =-630~\text{GeV}, M_{A} =141~\text{GeV}, \tan\beta=11~.
\eeq

With {\it FeynHiggs}, we compute the corresponding $R$-values and Higgs boson masses of benchmark point in
Eq.\ref{mhhh} and Eq.\ref{rhhh}
\beq
M_{h}=20.7~\text{GeV}, M_{H} = 123.3~\text{GeV}~,
\label{mhhh}
\eeq
\beq
R_{\gamma\gamma}=1.34, R_{ZZ}= R_{WW}=1.74, R_{b\bar{b}}=0.038, R_{\tau^{+}\tau^{-}}=0.72~.
\label{rhhh}
\eeq
At hadron colliders, the leading three production channels of $H$ are via gluon fusion, weak boson fusion as well as associated production.
\beq
pp\to H, jj H, WH, ZH
\eeq
$H$ can decay into $ZZ^{*}\to 4\ell$, $\gamma\gamma$, $\tau^{+}\tau^{-}$ and $WW^{*}\to 2\ell 2\nu$ just as
SM Higgs boson does. The corresponding events number in the final states respect to the SM Higgs boson prediction are given in Eq.\ref{rhhh}. However, $H\to b\bar{b}$ in our benchmark scenario is highly suppressed
with decay BR only 4\% of the SM prediction.  On the other hand, $H$ has significant decay BR into $hh$ states
as
\beq
\text{BR}(H\to hh)=39.3\%~.
\eeq
At $\tan\beta\sim 10$, {\it vev} of a $H_{d}$-like $h$ is small thus the coupling $hWW$.
The reduction in $hWW$ also results in partial width $\Gamma(h\to\gamma\gamma)$ is much smaller than the SM value.
$h$ dominantly decay into $b\bar{b}$ and $\tau^{+}\tau^{-}$ states as
\beq
\text{BR}(h\to b\bar{b})=85.8\%, \text{BR}(h\to \tau^{+}\tau^{-})=13.6\%~.
\eeq
Then $H$ can decay into $4b$, $4\tau^{\pm}$ or $2b2\tau^{\pm}$  with corresponding BR shown in parenthesis
\beq
 H\to h h\to b\bar{b}b\bar{b}~~(28.9\%), b\bar{b}\tau^{+}\tau^{-}~~(4.6\%), \tau^{+}\tau^{-}\tau^{+}\tau^{-}~~(0.73\%)~.
\eeq
Search of Higgs boson into $4b$, $2b2\tau^{\pm}$ and $4\tau^{\pm}$ final states have been
discussed in context of NMSSM \cite{jack, guiyu} as for $h\to a a$.
\cite{jack} studied Higgs boson $h$ from gluon fusion and Weak boson fusion production
with exactly the same final state as in our case. It typically requires 14~TeV LHC with 300~fb$^{-1}$ to
claim a 3-5$\sigma$ discovery due to large SM background.
To improve the signal over background ratio,
\cite{guiyu} focuses on the search of $h\to a a$ through associated production $Wh$/$Zh$ and
can reduce the required data to 100~fb$^{-1}$.
For the benchmark point in this paper, we have the associated production at 14~TeV LHC as
\beq
\sigma(pp\to W H) =1.59~\text{pb}, ~~\sigma(pp\to ZH)=0.94~\text{pb}~~(\text{14~TeV})
\eeq
the gluon fusion production rate for 8~TeV and 14~TeV LHC as
\beq
\sigma(gg\to H)(\text{8~TeV}) =22.18~\text{pb}, ~~\sigma(gg\to H)(\text{14~TeV}) =56.31~\text{pb}~.
\eeq
We estimate our signal rates, for instance, $\ell\nu +4b$ or $ \ell \nu +2 b 2\tau^{\pm}$ without any cut,
\bea
\sigma(pp\to WH\to \ell\nu +b\bar{b}b\bar{b} )& = & 102.5~\text{fb}\nonumber\\
\sigma(pp\to WH \ell \nu +b\bar{b} \tau^{+}\tau^{-}) & = & 16.25~\text{fb}
\eea
which is about 30\% less than the benchmark points of 120~GeV Higgs studied in \cite{guiyu}.
We argue that our benchmark point may require a little more data than claim in \cite{guiyu}.

For gluon fusion production, we want to point out that one particular interesting final states may
help the search. If $H$ is produced via gluon fusion and decay into four tau final states, one
can choose the final states as two same-sign leptonic taus with two hadronic taus
\beq
gg\to H\to h h\to \tau^{+}\tau^{-}\tau^{+}\tau^{-}\to \tau^{\pm}_{\ell}\tau^{\pm}_{\ell}\tau_{h}\tau_{h}
\eeq
which corresponds to same-sign di-lepton with one hadronic tau tag. The production rate
without any tag efficiency or kinematic cut is then
\beq
56.31~\text{pb}\times0.73\%\times 35\% \times 35\%\times 65\%\times 65\% \times 2\simeq 42.6~\text{fb}
\eeq
which made the final state also possible to search at 100~fb$^{-1}$.
On the other hand, this study requires much more realistic simulation including detector effects and polarized tau decay
treatment before drawing more convincing conclusions. We leave this study to experimental colleagues.

In summary, we discuss the possibility of $b\bar{b}$ final states with largest uncertainty being replaced by another final state.
In such scenarios, the Higgs signal shows
highly suppressed $b\bar{b}$, slightly reduced $\tau^{+}\tau^{-}$ and moderately enhanced gauge bosons comparing with the SM predictions. The model requires two different sources of electroweak symmetry breaking and radiative correction to $m_{b}$ strongly enhanced. However, large reduction in $b\bar{b}$ usually results large enhancement in $\tau^{+}\tau^{-}$ mode in particular. The reduction of $\tau^{+}\tau^{-}$ therefore implies that a new decay mode is inevitable. We find that a non-decoupling MSSM Higgs decay into lighter Higgs $H\to hh$ may fit the signature. Here, MSSM $H$ is identified as the 126~GeV resonance while $h$ is below $M_{H}/2$ and can evade the direct search bound at LEPII and Tevatron. However, the
scenario can only be realized in highly fine-tuned parameter region where $G_{Hhh}$ is tiny. Therefore, we argue the highly
suppressed $b\bar{b}$ is not likely.
Nevertheless, we in the end discuss the discovery potential of $H\to hh$ at the LHC which would require more than 14~TeV LHC with more than 100~fb$^{-1}$ of data to see any over three sigma deviation.

\section*{Acknowledgement}
ML is supported by the National Science Foundation of China (11135006) and National Basic Research Program of China (2010CB833000). KW is supported in part, by the Zhejiang University Fundamental Research Funds for the Central Universities (2011QNA3017) and the National Science Foundation of China (11245002,11275168). GZ is supported by the National Science Foundation of China (11075139).


\begin{thebibliography}{References}

\bibitem{today}
%\cite{:2012gk}%\bibitem{:2012gk}
  G.~Aad {\it et al.}  [ATLAS Collaboration],
  %``Observation of a new particle in the search for the Standard Model Higgs boson with the ATLAS detector at the LHC,''
  Phys.\ Lett.\ B
  [arXiv:1207.7214 [hep-ex]].
  %%CITATION = ARXIV:1207.7214;%%
%\cite{:2012gu}
%\bibitem{:2012gu}
  S.~Chatrchyan {\it et al.}  [CMS Collaboration],
  %``Observation of a new boson at a mass of 126 GeV with the CMS experiment at the LHC,''
  Phys.\ Lett.\ B
  [arXiv:1207.7235 [hep-ex]].
  %%CITATION = ARXIV:1207.7235;%%
    \bibitem{ww}
  ATLAS Collaboration, ATLAS-CONF-2012-098

\bibitem{Atlas-Spin}
ATLAS Collaboration, ATLAS-CONF-2012-169

\bibitem{CMS-Spin}
CMS Collaboration, CMS-PAS-HIG-12-041

  %\cite{Carena:1995bx}
\bibitem{Carena:1995bx}
  M.~S.~Carena, J.~R.~Espinosa, M.~Quiros and C.~E.~M.~Wagner,
  %``Analytical expressions for radiatively corrected Higgs masses and couplings in the MSSM,''
  Phys.\ Lett.\ B {\bf 355}, 209 (1995)
  [hep-ph/9504316].
  %%CITATION = HEP-PH/9504316;%%
  M.~S.~Carena, M.~Quiros and C.~E.~M.~Wagner,
  %``Effective potential methods and the Higgs mass spectrum in the MSSM,''
  Nucl.\ Phys.\ B {\bf 461}, 407 (1996)
  [hep-ph/9508343].
  %%CITATION = HEP-PH/9508343;%%


  %\cite{Ellwanger:2003jt}
\bibitem{jack}
  U.~Ellwanger, J.~F.~Gunion, C.~Hugonie and S.~Moretti,
  %``Towards a no lose theorem for NMSSM Higgs discovery at the LHC,''
  hep-ph/0305109.
  %%CITATION = HEP-PH/0305109;%%
  %\cite{Ellwanger:2005uu}
%\bibitem{Ellwanger:2005uu}
  U.~Ellwanger, J.~F.~Gunion and C.~Hugonie,
  %``Difficult scenarios for NMSSM Higgs discovery at the LHC,''
  JHEP {\bf 0507}, 041 (2005)
  [hep-ph/0503203].
  %%CITATION = HEP-PH/0503203;%%
  %\cite{Chang:2005ht}
%\bibitem{Chang:2005ht}
  S.~Chang, P.~J.~Fox and N.~Weiner,
  %``Naturalness and Higgs decays in the MSSM with a singlet,''
  JHEP {\bf 0608}, 068 (2006)
  [hep-ph/0511250].
  %%CITATION = HEP-PH/0511250;%%

%\cite{Carena:2007jk}
\bibitem{guiyu}
  M.~Carena, T.~Han, G.~-Y.~Huang and C.~E.~M.~Wagner,
  %``Higgs Signal for h $\to$ aa at Hadron Colliders,''
  JHEP {\bf 0804}, 092 (2008)
  [arXiv:0712.2466 [hep-ph]].
  %%CITATION = ARXIV:0712.2466;%%



%\cite{Williams:2011bu}
\bibitem{higherorder}
  K.~E.~Williams, H.~Rzehak and G.~Weiglein,
  %``Higher order corrections to Higgs boson decays in the MSSM with complex parameters,''
  Eur.\ Phys.\ J.\ C {\bf 71}, 1669 (2011)
  [arXiv:1103.1335 [hep-ph]].
  %%CITATION = ARXIV:1103.1335;%%

%\cite{Belyaev:2006rf}
\bibitem{cp}
  A.~Belyaev, Q.~-H.~Cao, D.~Nomura, K.~Tobe and C.~-P.~Yuan,
  %``Light MSSM Higgs boson scenario and its test at hadron colliders,''
  Phys.\ Rev.\ Lett.\  {\bf 100}, 061801 (2008)
  [hep-ph/0609079].
  %%CITATION = HEP-PH/0609079;%%

%\cite{Heinemeyer:2011aa}
\bibitem{Heinemeyer2011}
  S.~Heinemeyer, O.~Stal and G.~Weiglein,
  %``Interpreting the LHC Higgs Search Results in the MSSM,''
  Phys.\ Lett.\ B {\bf 710}, 201 (2012)
  [arXiv:1112.3026 [hep-ph]].
  %%CITATION = ARXIV:1112.3026;%%


%\cite{Bottino:2011xv}
\bibitem{Bottino}
  A.~Bottino, N.~Fornengo and S.~Scopel,
  %``Phenomenology of light neutralinos in view of recent results at the CERN Large Hadron Collider,''
  Phys.\ Rev.\ D {\bf 85}, 095013 (2012)
  [arXiv:1112.5666 [hep-ph]].
  %%CITATION = ARXIV:1112.5666;%%

%\cite{Christensen:2012ei}
\bibitem{taohan}
  N.~D.~Christensen, T.~Han and S.~Su,
  %``MSSM Higgs Bosons at The LHC,''
  Phys.\ Rev.\ D {\bf 85}, 115018 (2012)
  [arXiv:1203.3207 [hep-ph]].
  %%CITATION = ARXIV:1203.3207;%%

%\cite{Hagiwara:2012mg}
\bibitem{hagiwara}
  K.~Hagiwara, J.~S.~Lee and J.~Nakamura,
  %``Properties of 126 GeV Higgs boson in non-decoupling MSSM scenarios,''
  arXiv:1207.0802 [hep-ph].
  %%CITATION = ARXIV:1207.0802;%%

  %\cite{Benbrik:2012rm}
\bibitem{heinemeyer}
  R.~Benbrik, M.~G.~Bock, S.~Heinemeyer, O.~Stal, G.~Weiglein and L.~Zeune,
  %``Confronting the MSSM and the NMSSM with the Discovery of a Signal in the two Photon Channel at the LHC,''
  arXiv:1207.1096 [hep-ph].
  %%CITATION = ARXIV:1207.1096;%%

%\cite{Arbey:2012dq}
\bibitem{Arbey}
  A.~Arbey, M.~Battaglia, A.~Djouadi and F.~Mahmoudi,
  %``The Higgs sector of the phenomenological MSSM in the light of the Higgs boson discovery,''
  JHEP {\bf 1209}, 107 (2012)
  [arXiv:1207.1348 [hep-ph]].
  %%CITATION = ARXIV:1207.1348;%%

%\cite{Belanger:2012tt}
\bibitem{Belanger:2012tt}
  G.~Belanger, U.~Ellwanger, J.~F.~Gunion, Y.~Jiang, S.~Kraml and J.~H.~Schwarz,
  %``Higgs Bosons at 98 and 126 GeV at LEP and the LHC,''
  arXiv:1210.1976 [hep-ph].
  %%CITATION = ARXIV:1210.1976;%%

%\cite{Drees:2012fb}
\bibitem{Drees}
  M.~Drees,
  %``A Supersymmetric Explanation of the Excess of Higgs--Like Events at the LHC and at LEP,''
  arXiv:1210.6507 [hep-ph].
  %%CITATION = ARXIV:1210.6507;%%
  %\cite{Ke:2012zq}
%\cite{Ke:2012zq}
\bibitem{kejiwei}
  J.~Ke, H.~Luo, M.~-x.~Luo, K.~Wang, L.~Wang and G.~Zhu,
  %``Revisit to Non-decoupling MSSM,''
  arXiv:1211.2427 [hep-ph].
  %%CITATION = ARXIV:1211.2427;%%
%\cite{Bechtle:2012jw}
\bibitem{Bechtle:2012jw}
  P.~Bechtle, S.~Heinemeyer, O.~St [] l, T.~Stefaniak, G.~Weiglein and L.~Zeune,
  %``MSSM Interpretations of the LHC Discovery: Light or Heavy Higgs?,''
  arXiv:1211.1955 [hep-ph].
  %%CITATION = ARXIV:1211.1955;%%





%\cite{Frank:2006yh}
\bibitem{feynhiggs}
  M.~Frank, T.~Hahn, S.~Heinemeyer, W.~Hollik, H.~Rzehak and G.~Weiglein,
  %``The Higgs Boson Masses and Mixings of the Complex MSSM in the Feynman-Diagrammatic Approach,''
  JHEP {\bf 0702}, 047 (2007)
  [hep-ph/0611326].
  %%CITATION = HEP-PH/0611326;%%
%\cite{Degrassi:2002fi}
%\bibitem{Degrassi:2002fi}
  G.~Degrassi, S.~Heinemeyer, W.~Hollik, P.~Slavich and G.~Weiglein,
  %``Towards high precision predictions for the MSSM Higgs sector,''
  Eur.\ Phys.\ J.\ C {\bf 28}, 133 (2003)
  [hep-ph/0212020].
  %%CITATION = HEP-PH/0212020;%%
  %\cite{Heinemeyer:1998np}
%\bibitem{Heinemeyer:1998np}
  S.~Heinemeyer, W.~Hollik and G.~Weiglein,
  %``The Masses of the neutral CP - even Higgs bosons in the MSSM: Accurate analysis at the two loop level,''
  Eur.\ Phys.\ J.\ C {\bf 9}, 343 (1999)
  [hep-ph/9812472].
  %%CITATION = HEP-PH/9812472;%%
%\cite{Heinemeyer:1998yj}
%\bibitem{Heinemeyer:1998yj}
  S.~Heinemeyer, W.~Hollik and G.~Weiglein,
  %``FeynHiggs: A Program for the calculation of the masses of the neutral CP even Higgs bosons in the MSSM,''
  Comput.\ Phys.\ Commun.\  {\bf 124}, 76 (2000)
  [hep-ph/9812320].
  %%CITATION = HEP-PH/9812320;%%


%\cite{Bechtle:2008jh}
\bibitem{higgsbounds}
  P.~Bechtle, O.~Brein, S.~Heinemeyer, G.~Weiglein and K.~E.~Williams,
  %``HiggsBounds: Confronting Arbitrary Higgs Sectors with Exclusion Bounds from LEP and the Tevatron,''
  Comput.\ Phys.\ Commun.\  {\bf 181}, 138 (2010)
  [arXiv:0811.4169 [hep-ph]].
  %%CITATION = ARXIV:0811.4169;%%
  P.~Bechtle, O.~Brein, S.~Heinemeyer, G.~Weiglein and K.~E.~Williams,
  %``HiggsBounds 2.0.0: Confronting Neutral and Charged Higgs Sector Predictions with Exclusion Bounds from LEP and the Tevatron,''
  Comput.\ Phys.\ Commun.\  {\bf 182}, 2605 (2011)
  [arXiv:1102.1898 [hep-ph]].
  %%CITATION = ARXIV:1102.1898;%%




%\cite{Crivellin:2012jv}
\bibitem{Crivellin:2012jv}
  A.~Crivellin, J.~Rosiek, P.~H.~Chankowski, A.~Dedes, S.~Jaeger and P.~Tanedo,
  %``SUSy_FLAVOR v2: A Computational tool for FCNC and CP-violating processes in the MSSM,''
  arXiv:1203.5023 [hep-ph].
  %%CITATION = ARXIV:1203.5023;%%



  %\cite{Carena:2011aa}
\bibitem{carlos}
  M.~Carena, S.~Gori, N.~R.~Shah and C.~E.~M.~Wagner,
  %``A 126 GeV SM-like Higgs in the MSSM and the $\gamma \gamma$ rate,''
  JHEP {\bf 1203}, 014 (2012)
  [arXiv:1112.3336 [hep-ph]].
  %%CITATION = ARXIV:1112.3336;%%
%\cite{Carena:2012gp}
%\bibitem{Carena:2012gp}
  M.~Carena, S.~Gori, N.~R.~Shah, C.~E.~M.~Wagner and L.~-T.~Wang,
  %``Light Stau Phenomenology and the Higgs \gamma\gamma Rate,''
  JHEP {\bf 1207}, 175 (2012)
  [arXiv:1205.5842 [hep-ph]].
  %%CITATION = ARXIV:1205.5842;%%
%\cite{Ke:2012qc}
%\bibitem{Ke:2012qc}
  J.~Ke, M.~-X.~Luo, L.~-Y.~Shan, K.~Wang and L.~Wang,
  %``Searching SUSY Leptonic Partner at the CERN LHC,''
  arXiv:1207.0990 [hep-ph].
  %%CITATION = ARXIV:1207.0990;%%
  %\cite{Buckley:2012em}







%\cite{Djouadi:2005gj}
\bibitem{physicsreport}
%\cite{Haber:1996fp}
%\bibitem{Haber:1996fp}
  H.~E.~Haber, R.~Hempfling and A.~H.~Hoang,
  %``Approximating the radiatively corrected Higgs mass in the minimal supersymmetric model,''
  Z.\ Phys.\ C {\bf 75}, 539 (1997)
  [hep-ph/9609331].
  %%CITATION = HEP-PH/9609331;%%
  A.~Djouadi,
  %``The Anatomy of electro-weak symmetry breaking. II. The Higgs bosons in the minimal supersymmetric model,''
  Phys.\ Rept.\  {\bf 459}, 1 (2008)
  [hep-ph/0503173].
  %%CITATION = HEP-PH/0503173;%%




\end{thebibliography}
\end{document}